# Formation of non-collinear magnetic states in the $Fe_{2-x}Mn_xAs$ system

Val`kov V.I.[1]., Golovchan A.V.[1,2]., Varyukhin D.V.[1], Szymczak H.[3]., Dyakonov V.P.[1,3]

[1]Donetsk Institute for Physics and Engineering named after A.A. Galkin NAS of Ukraine, 72 R. Luxemburg Str., 83114, Donetsk, Ukraine

[2]Donetsk National University, 24 Universitetskaya Str., 83055, Donetsk, Ukraine

[3]Institute of Physics, PAS, 02-668 Warsaw, Al, Lotnikow 32/46, Poland

E-mail: valkov09@gmail.com

**Abstract**

Within the frameworks of the model using *ab initio* calculation data on d-band filling and on the shape of density of d-electron states in $Fe_{2-x}Mn_xAs$, the mechanisms of stabilization of non-collinear magnetic ordered phases observed within the range of $1.19 \leq x \leq 1.365$ are considered. The relation between baric peculiarities of the changes in stability of noncollinear phases and changes in the electronic structure characteristics of $Fe_{2-x}Mn_xAs$ due to lattice deformation is disclosed.

It is found that stabilization of non-collinear phases and change in type of order-order phase transition with decreasing manganese is determined by competition between the electronic filling of the d-band, shape of the d-electron density of states and structural features of these phases.

## 1. Introduction

The $Fe_{2-x}Mn_xAs$ system occupies a special position among series of the antiferromagnetic(AF) pnictides of transition metals. This fact is determined by the existing narrow range of manganese content ($x = 1.19 \div 1.365$) where the order-order magnetic phase transition below Neel temperature takes place. As a result, samples obtain spontaneous magnetization [1, 2], Fig. 1. This result violates the intuitive understanding of formation of magnetically ordered phases by mixing of two antiferromagnetic compounds. The emergence of an uncompensated magnetic moment may be caused by competition between the spin polarization of d-states of the atoms of manganese and iron, chaotically located in tetrahedral positions. This competition is the reason for the formation of the low-temperature low-ferrimagnetic phase (LFi$_1$) as a noncollinear (canted) structure [3].

On the one hand, the canted ferrimagnetic phase LFi$_1$ keeps antiferromagnetic characteristics of the original antiferromagnetic compounds Fe$_2$As, MnFeAs, Mn$_2$As; on the other hand, there is an uncompensated macroscopic magnetic moment that characterizes ferrimagnetic state.

In our previous paper [4], the electronic structures of the collinear magnetic states of Fe$_{0.69}$Mn$_{1.31}$As for different types of hydrostatic and uniaxial pressures were calculated from the first principles. At the same time, within the model frameworks, qualitative analysis of the cascade of magnetic-field induced phase transitions AF-LFi$_1$-LFi$_2$ was carried out.

The peculiarity of the present work is not only the analysis of conditions of appearance of the canted LFi$_1$, LFi$_2$ phases in a limited concentration range of Mn ($1.19 \leq x \leq 1.365$), but also explanation of the reduction of the spontaneous magnetization and the changes of type of the order-order phase transitions from the first to the second one with increasing manganese concentration within this range. Within the frameworks of the model, the relationship between specific behavior of canted phases under pressure and the changes of the characteristics of the electronic structure of Fe$_{2-x}$Mn$_x$As due to baric deformations of the lattice is disclosed.

Here, in contrast to the dimensionless results of [4], the results of the present work are evaluated in real units of measurement of magnetic induction (*B*), magnetic moment (*M*) and energy (*E*).

## 2. **Experimental data**

Now we present the most important facts for our consideration taken from the available literature data:

1. The range of the existence of spontaneous low-temperature low-ferrimagnetic (LFi$_1$) phase is limited by the concentration range of $1.19 \leq x \leq 1.365$ [1, 2], Fig. 1.

2. The order of spontaneous and magnetic-field-induced transition from antiferromagnetic to low-ferrimagnetic state at opposite ends of the considered range of manganese concentrations is different [2].

3. There exist reversible magnetic-field induced phase transitions between the canted phases LFi$_1$-LFi$_2$ [5, 6, 7], which are accompanied by a significant increase in the magnetization of the samples.

4. Displacement of boundaries of spontaneous and induced phase transitions AF-LFi$_1$ has some peculiarities under the different types of uniaxial pressure [8], Fig. 3.

Figure 2 shows that with increase in concentration of manganese in the range of existence of LFi$_1$ phase for the Fe$_{2.1-x}$Mn$_x$As(a) and Fe$_{2-x}$Mn$_x$As (b) systems, spontaneous magnetic moment (m) of the low-ferrimagnetic phase is reduced, and the values of magnetic moment of the LFi$_2$ phase induced by high magnetic field are increased. Here, according to Fig. 2c, M does not reach the values of the paramagnetic moment ($\mu_{PM}$) and the saturation magnetic moment of collinear ferromagnetic (M$_{FM}$) and ferrimagnetic (M$_{FIM-I}$) phases. We should also note that in the samples related to the beginning of the range (left part), the AF-LFi$_1$ phase transitions are accompanied by sharp changes of volume and lattice parameters *c* and *a* [2, 9], i.e. they are the first-order phase transitions. Outside the left edge of the concentration range, this fact results in irreversible stepwise occurrence of the LFi$_1$ state, as magnetic field increases [5]. On the contrary, at the end of range (right part) these quantities vary smoothly [2].

Investigation of stability of the LFi$_1$ phase with respect to the type of uniaxial compression (Fig. 3) revealed that compression of a monocrystalline sample perpendicular to the tetragonal *c* axis is a stabilizing factor. According to Fig. 3a, in this case, pressure increase results in shift of magnetic field induced transitions toward the region of lower fields. Conversely the effect of uniaxial pressure along the *c* axis (Fig. 3b) shifts the magnetization curve toward higher fields stronger than analogous changes in the case of hydrostatic pressure. General conception of the shifts of phase boundaries in the temperature-pressure plane (T–P) is presented in [8] for various types of pressure (Fig. 4).

## 3. DFT calculations

For *ab initio* calculations, we used a fully relativistic KKR-CPA method (SPRKKR package [10]). Calculation of electronic structure was carried out in the coherent potential approximation for the model of a disordered alloy: we assumed that the atoms of Fe$_I$ and Mn$_I$ are randomly distributed over tetrahedral positions. The crystal lattice parameters for the Fe$_{2-x}$Mn$_x$As system are taken from [11]. Reported in [4, 12], previous *ab initio* calculations of the electronic structure of collinear magnetically ordered states indicate that, for example, the antiferromagnetic state AF1-1 has the lowest energy at *x*=1.29 (E(AF1-1)-E(FM) = –0.00309 Ry). The energy of the nearest state with spontaneous magnetization, i.e. the ferrimagnetic FIM-I state, is much higher (E(FIM-I)-E(FM) = –0.0006 Ry). Thus, the occurrence of spontaneous magnetization in Fe$_{2-x}$Mn$_x$As is most likely caused by stabilization of a noncollinear ferrimagnetic state in a narrow interval of manganese concentration (1.19 ≤ *x* ≤ 1.365). It is assumed that the gain in energy of the low-ferrimagnetic state LFi$_1$ compared to AF1-1 arises from the competition of the kinetic energy of d-electrons (it depends on the number of electrons and shape of the density of electronic states) and the exchange energy (it depends only on the number of interacting d-electrons in the band). However, since straight calculations of noncollinear structures in the Fe$_{2-x}$Mn$_x$As system are time consuming at present, we will use the model [6] for interpretation of the experimental data. The model uses information about the number of d-electrons and the shape of the density of electronic states in the nonmagnetic phase.

Figure 5 presents energy dependences of the density of d-electron states for the nonmagnetic phase $DOS_{dNM}(E)$ in the range of $1.15 \leq x \leq 1.45$. Peculiarities of structure evolution depending on manganese content are shown in Figure 6. It also illustrates coordinate dependence of the number of d-electrons per a state $n(x)=N_d(x)/20$, which allows us to identify a particular compound by the electronic filling of the d-band. $N_d(x)$ was calculated by the formula

$$N_d(x) = [2N_d(Mn_I) \cdot (x-1) + 2N_d(Fe_I) \cdot (2-x)(Fe_I) + 2N_d(Mn_{II}))], \qquad (1)$$

where the values $N_d(Mn_I)$, $N_d(Fe_I)$, $N_d(Mn_{II})$ were calculated in the present work for selected lattice parameters [11].

We suppose that the degree of filling of magnetoactive band in the non-magnetic phase ($N_d^{NM}$) and the shape of the density of electronic states $DOS_{dNM}(E)$ are responsible for the further formation of spontaneous and magnetic-field-induced canted phases in the range of $1.19 \leq x \leq 1.365$. Changing the above characteristics with the Mn content variation results in a change in the stability and structural features of canted phases found in [3]. The latter is the angle ($\Theta$) between the ferromagnetic and antiferromagnetic components of the total magnetic moment of a cell.

As seen in Figure 5 and Figure 6, $DOS_{dNM}(E)$ is changed with increasing $x$; the area of occupied states ($\Delta E_{filled}$) becomes narrowed on the background of total narrowing of the d-band ($\Delta E$), and the area of empty states ($\Delta E_{empty}$) widens. These changes are accompanied by an increase in heights of the second ($D_2$) and the fifth ($D_5$) peaks of $DOS_{NM}(E)$ and reduction of the total number of d-electrons. In Figures 5 and 6, as an effective width of occupied and empty states of $DOS_{NM}(E)$, the energy range ($\Delta E_{filled}= E_F–E_L$) and ($\Delta E_{empty}= E_R–E_F$) are chosen, which are measured at one third height of $DOS_{NM}(E)$. Intuitively, such changes in the electronic characteristics are quite explainable: substitution of the part of iron atoms (with ionic radius of 0.76Å) by manganese atoms with large ionic radius (0.8Å) and a smaller number of 3d-electrons should result in a decrease in $N_d$, increase of the unit-cell volume, and succeeding narrowing of d-band [13, 14].

Based on the experimental results, we can conclude that the combination of the above parameters in samples of $Fe_{2-x}Mn_xAs$ ($1.19 \leq x \leq 1.365$) should lead to the existence of spontaneous and induced by magnetic field low-ferrimagnetic phases $LFi_1$, $LFi_2$. On the other hand, for samples with $x<1.19$ and $x>1.365$, the values of these parameters make spontaneous appearance of canted phase $LFi_1$ energetically unfavorable (compared to AF phase).

The *ab initio* calculations of spin-polarized electronic structure of $Fe_{0.69}Mn_{1.31}As$ for different compression of the unit cell [4] allow us to build a similar dependences of the parameters of electronic structure of non-magnetic phase as functions of relative volume deformations $\omega = (V-V_0)/V_0$ (Fig. 7).

It is assumed that the hydrostatic pressure corresponds to a proportional decrease in unit cell parameters. Compression along the tetragonal axis (P||c) was modeled in such a way that the reduction of *c*-axis was accompanied by an increase of *a*-axis with a resulting volume reduction. Uniaxial pressure in the basal plane (P⊥*c*) was modeled as a uniform reduction of the parameter *a* with an increase in parameter *c* and the overall increase in unit cell volume. In this case, we can highlight the most common patterns of change in the structure of nonmagnetic density of electronic states and the degree of electronic filling as reaction on the corresponding types of deformation. A uniform or hydrostatic compression (*c*/*a*=const, $\omega<0$) slightly reduces d-band filling $n(N_d)$, increases the parameters $\Delta E_{filled}$, $\Delta E_{empty}$, effective d-band width ($\Delta E$) and reduces $D_2$, $D_5$ peak heights (Fig. 7a). Compression along the *c*-axis results in stronger reduction of d-band; it narrows the area of the filled states ($d\Delta E_{filled}/d|\omega| < 0$) and extends the area of the empty states ($d\Delta E_{empty}/d|\omega| > 0$), with overall increase of band width $\Delta E$, Fig. 7b. Uniaxial strain (P⊥c), which increases the unit-cell volume ($\omega>0$), narrows the region of empty states ($d\Delta E_{empty}/d\omega<0$) with increasing width of the area of occupied states ($d\Delta E_{filled}/d\omega>0$), so the effective width of the d-band is narrowed ($d\Delta E/d\omega<0$). It is also accompanied by an increase in the $D_2$, $D_5$ peak heights, Fig. 7c. In contrast to the uniform compression, both types of uniaxial compression did not significantly change the electronic d-band filling (*n*).

Thus, changes in selected parameters of the electronic system due to variations in chemical composition and interatomic distances may form the mechanism of change in stability of magnetic phases. Peculiarities of occurrence of each of the two factors (the shape of the density of electronic states and the number of d-electrons) can be conveniently analyzed on a qualitative model. Here we use two-site model of itinerant electrons.

### 4. Two-site model of itinerant-electrons

In this model approach [4], each unit cell of the original $Fe_{2-x}Mn_xAs$ containing two formula units (4 magnetic atoms with 20 itinerant d-states) is associated with diatomic cell *j* with two s-like d-states. Then, in the model unit cell including two formally different sites (*a,b*), ferromagnetic and antiferromagnetic polarization of electronic spectrum can be described by spatially homogeneous irreducible vectors of ferromagnetism $M$ and antiferromagnetism $L$, which act as magnetic order parameters. Determining relative orientation of magnetic moments in the positions *a* and *b*, vectors $M = \langle \hat{F} \rangle$ and $L = \langle \hat{L} \rangle$ correspond to the statistical average of operators $\hat{F}_j$ and $\hat{L}_j$ with model Hamiltonian (A1). Within the functional integral theory and single-site (molecular field) approximation [4, 14] for rigidly-directed exchange fields $\vec{\xi},\vec{\eta}$, free energy $\mathbf{F}(\xi,\eta)$ is calculated and equations of state $\partial \mathbf{F}/\partial \xi = 0$, $\partial \mathbf{F}/\partial \eta = 0$ (A11) are solved. The resulting solutions $m_0(B)=g\mu_B M$, $m_Q(B)=g\mu_B L$ are compared with experimental magnetization curves of the system under consideration. Here $B$ is external magnetic field ($B = \mu_0 H_0$). Solutions of the system (A11) are sought at constant intra-atomic exchange integral $J$, which is the same for all investigated alloys; the number of electrons $n(x, P_i)$ and bare functions $g_{00}(\varepsilon, P_i)$, $g_{11}(\varepsilon, P_i)$ depend on composition and pressure $P_i$; Fig. 8a.

The value of intra-atomic exchange integral $J$ is estimated from the condition $J\mu=\Delta E_{EX}$ where μ is magnetic moment per d-state, and $\Delta E_{EX}$ is exchange splitting, which is defined as the displacement of centers of d-electron density of "spin up" states $DOS_{up}(E)$ and "spin down" ones $DOS_{down}(E)$ measured at a height of 1/3 of each function in the ferromagnetic phase. In the interval of $1.15 \leq x \leq 1.4$, $J$ does not exceed 0.378Ry. In model calculations, the value 0.349Ry was used. It was assumed that the model density of electronic states $G_{NM}(E)$ (Fig. 8b) can adequately describe the field dependence of $\sigma(H)$ (Fig. 3) and $M(B)$ (Fig. 11), if its shape is similar to the shape of density of d-states $DOS_{dNM}(E)$, obtained from *ab initio* calculations for the selected compounds, for example $Fe_{0.69}Mn_{1.31}As$. In our model, this assumption is realized by choosing the bare functions $g_{00}(\varepsilon)$, $g_{11}(\varepsilon)$, Fig. 8a, (A12).

## 5. Results

In the investigated range of manganese content ($1.19 \leq x \leq 1.365$) change in $x$ at constant volume ($\omega = 0$) or volume ($|\omega| \neq 0$) at constant $x$ is simulated by corresponding change of the number of electrons $n(x)$ ($1.224 < n < 1.206$)(1), angle $\Theta$ and parameters $\varepsilon_L$, $\varepsilon_R$, $g_2$, $g_4$, $g_5$ of function $g_{00}(\varepsilon)$, (Fig. 8a).

At atmospheric pressure, for a given $J$, $n$ and bare functions $g_{00}(\varepsilon)$ and $g_{11}(\varepsilon)$, there are two types of solutions that determine the appropriate minimum of the free energy in zero field (Fig. 10), as well as spontaneous and field-induced states (Fig. 11). The values marked in the figure as $m_{FM0}$, are the ferromagnetic ($\eta\equiv 0$) solutions of (A11) in zero-field and exist for any $n(x)$. These solutions correspond to the minimum of free energy $\mathbf{F}1(m_0)\equiv \mathbf{F}(m_0,m_Q\equiv 0)$, and define a metastable ferromagnetic state Fm, which can be compared with the FM state as calculated from first principles (Fig. 2, $M_{FM}(x)$). The second solution is indicated in Figure 10 as $m_{00}$. It is associated with the removal of constraint $\eta\equiv 0$ and corresponds to the minimum of free energy $\mathbf{F}2(m_0) \equiv \mathbf{F}(m_0, m_Q(m_0))$. This solution can describe either ferromagnetic state Fm in large fields at $m_0=m_{FM}$, $m_Q=0$ or canted states $LFm_1$, $LFm_2$ when $0<|m_0|<|m_{FM}|$, $m_Q>0$ (Fig. 11a). The latter states can be associated with the experimentally observed weakly ferrimagnetic phases LFi$_1$, LFi$_2$. In Fig. 11a, for $n$=1.2246, the dependences $m_{FM}(B)$, $m_0(B)$ describe the sequence of induced transitions AF-LFm$_1$-LFm$_2$-Fm, the first of which simulates the experimentally observed irreversible phase transition AF-LFi$_1$ in Fe$_{1.95-x}$Mn$_x$As with $x$ = 1.185 [5]. According to this dependence, the increase in the ferromagnetic component $m_0$ is accompanied by a decrease in the antiferromagnetic $m_Q$ down to the complete vanishing in the fields $B >> B_{12}$, when the solutions $m_0(B)$ and $m_Q(B)$ merge. As can be seen in Figure 9, canted state LFm$_1$ with H$_0$ = 0 corresponds to the absolute energy minimum **F2**(m$_0$) only in a limited range of $n$ = 1.2245 ÷ 1.207, which unambiguously (1) specifies the interval of manganese concentrations $x$ = 1.19 ÷ 1.365. Outside this interval, when $n \geq 1.2245$ ($x \leq 1.19$) and $n < 1.207$ ($x > 1.365$), the minimum energy corresponds to AF phase ($m_{00}$=0, $m_{Q0} \neq 0$) and canted LFm$_1$ states can arise only in magnetic field at induced order-order phase transitions AF-LFm$_1$, Fig. 11. These transitions correspond to irreversible and reversible magnetic-field-induced transitions AF-LFi$_1$ observed in [7, 8]. It should be noted that the nature of magnetic-field-induced transitions AF-LFm$_1$ at the end points of the interval differs significantly. According to the experimental data [5], at $x \leq 1.19$ ($n \geq 1.2236$), induced transitions are magnetic first-order phase transitions with their characteristic field hysteresis and large jump in the magnetization (Fig. 11a). Near the right endpoint of the interval ($n < 1.207$, $x > 1.365$), the transition AF-LFm$_1$ (Fig. 11b) becomes closer to the second order transition in agreement with experimental data on the behavior of the Fe$_{a-x}$Mn$_x$As system at large $x$ [2, 9].

As mentioned above, the advantage of model approach is that you can estimate the contributions of various factors in the formation of canted states. Analysis has shown that canted state LFm$_1$ (identifiable with weakly ferrimagnetic phase LFi$_1$) exists due to distortion of the angular structure, which according to the model, should grow with increasing occupancy of d-states, accompanied by a reduction of manganese. In undistorted symmetric phase (angle $\Theta$ between ferromagnetic($m_{00}=m_0(H_0=0)$) and antiferromagnetic($m_{Q0}=m_Q(H_0=H_Q=0)$) components of the magnetic moment or components of the spin fields $\vec{\xi}, \vec{\eta}$ in (A5) is 90º($e_x$=1, ($\vec{\xi} \cdot \vec{\eta} = 0$)), the state LFm$_1$ is not realized. When deviation $\Theta$ from 90º, an additional contribution to the energy $(J)^2(\vec{\xi}\eta)^2$ (A5) starts. It results in an asymmetrical canted structure formed by ferromagnetic and antiferromagnetic components of magnetic moment (Fig. 12) and stabilizes the LFm$_1$ state. Figure 10 shows that the increase of $\Theta$, that accompanies the increase of $x$, (Fig. 9), leads to a shift of the energy minimum to smaller $m_{00}$. Such trends are qualitative differences of the character of LFm$_1$ destabilization at the endpoints of the interval $1.19 \leq x \leq 1.365$. With decreasing $x$, approach to the lability boundary of the LFm$_1$ state is not accompanied by so significant change in the spontaneous ferromagnetic components $m_{00}$, which is the result of competition of several factors: destabilizing increase of $n$ is partially compensated by increase (decrease) in $|\varepsilon_L|(\varepsilon_R)$ when the angle $\Theta$ is reduced. When moving along the $x$ right to other endpoint, the stabilizing influence of decreasing $n$ is compensated by a decrease (increase) $|\varepsilon_L|(\varepsilon_R)$ with increasing $\Theta$. In this case, the approximation to the lability boundary of LFm$_1$ state is accompanied by strong decrease of the ferromagnetic zero-field components ($dm_{00}/dx<0$) and the monotonic growth ($dm_{02AV2}/dx>0$) of its initial value $m_{02AV2}$ of LFm$_2$ state in induced magnetic field transition LFm$_1$-LFm$_2$ in the field B=B$_{AV2}$ where F2($m_{01AV2}$) = F1($m_{02AV2}$), (Fig. 10,11).

In another case ($d\Theta/dx\leq0$), the dependence $m_{00}(x)$ in the range of $1.19 \leq x \leq 1.365$ stops decrease and the transition to the AF state is not accompanied by a preliminary smooth decrease of $m_{00}(x)$. There will be also observed an increase of $m_{02,1}$ at induced transitions LFm$_1$-LFm$_2$.

At model description of pressure effects, we calculated the magnetization curves for the respective sets of parameters $n, \varepsilon_L, \varepsilon_R, g_2, g_4, g_5$ as functions of volumetric deformation ω of different types. Changes in these parameters were compared with those changes calculated *ab initio* $\Delta E_{filled}(\omega)$, $\Delta E_{empty}(\omega)$, $D_2(\omega)$, $D_5(\omega)$ for the Fe$_{0.69}$Mn$_{1.31}$As composition (Fig. 7). At the same time, due to the lack of experimental data, the value of $\Theta(\omega)$ was assumed constant (≈77°). The results of calculations are presented in Figure 13. As can be seen, the model curves qualitatively reproduce the effects of both comprehensive and uniaxial compression. In complete agreement with the experimental data, uniform compression and compression along the tetragonal axis have a destabilizing effect on LFm$_1$, LFm$_2$ (Fig. 13a, b). The quantitative difference between the effects of uniaxial ($\|c$) and comprehensive compressions appears in a larger (at equal relative deformations) displacement of spontaneous and magnetic-field-induced transitions AF-LFm$_1$-LFm$_2$ to higher fields. Hydrostatic compression ω=–0.8% (modeled by increasing of parameters $\varepsilon_R$, $|\varepsilon_L|$ and reduction of $g_2, g_4, g_5, n$, Fig. 13a) destabilizes broadening of empty states ($\Delta E_{empty}$) and is partially compensated by an increase in the occupied states width ($\Delta E_{filled}$) (Fig. 7a), which is a stabilizing factor. Compression along the tetragonal axis (ω=–0.6%) is modeled by more significant increase in $\varepsilon_R$ and reduction of the parameters n, $|\varepsilon_L|$ and $g_4$ (Fig. 8). The result is a significant shift of all states toward the region of higher magnetic fields. In this case, the magnetization curves (Fig. 13b) can be compared with the experimental ones (Fig. 3b), where the impact of uniaxial pressure along the tetragonal axis *c* shifts low-ferrimagnetic phase LFi$_1$ to higher magnetic fields. Unlike the two previous cases, the compression in the direction perpendicular to the tetragonal axis (ω=+0.588%) results in stabilization of LFm$_1$, LFm$_2$ states. It can be seen as increase in the spontaneous magnetic moment $m_{10}$, $m_{Fm}$ of LFm$_1$, Fm states and shift of the induced transitions LFm$_1$-LFm$_2$ to lower magnetic fields (Fig. 13c). Such behavior is consistent with experimental data [8] (Fig. 3a) and is modeled by decrease $\varepsilon_R$ and increase $\varepsilon_L$ with a decrease in their amount $d\varepsilon$ and constant $n$, in agreement with Fig. 7c. Such behavior is consistent with the values of $M_{FM}(\omega)$ calculated from the first principles (Fig. 6c) and experimental data [8] (Fig. 3a) and it can be modeled by $\varepsilon_R$ decrease and increase of $n$, $g_4$, and $\varepsilon_L$ at reduction of their amount $\Delta\varepsilon$.

**Conclusion**

1. Is established that stabilization of spontaneous canted states in the Fe$_{a-x}$Mn$_x$As system is energetically favorable at non-orthogonal arrangement of ferromagnetic and antiferromagnetic components of the total magnetic moment.

2. Reduction of energy due to the disappearance of the spontaneous magnetization and the stabilization of the antiferromagnetic phase at the endpoints of the interval ($1.19 \leq x \leq 1.365$) result from changes in the occupancy of d-bands and structural elements of the shape of the density of electronic states.

3. Change in the character of order-order phase transitions in the interval of $1.19 \leq x \leq 1.365$ is determined by increasing trend to orthogonal arrangement of ferromagnetic and antiferromagnetic components of magnetic moment.

4. Peculiarities in the behavior of spontaneous and magnetic-field-induced order-order phase transitions under pressure are associated with the character of the barometric renormalization of d-bands occupancy and shape parameters of the density of electronic states.

This work was supported by the State Fund for Fundamental Researches of Ukraine (Project No. 41.1/038) and by European Fund for Regional Development (Contract No. UDA-POIG.01.03.01-00-058/08/00). The calculations have been performed at the grid-cluster of Donetsk Institute for Physics and Engineering NASU under support of grant No. 232.

**Appendix**

In the present approach, the Hamiltonian of model system (A1) with two s-like d-states per cell j contains three terms that describe the character of d-band states (A2a), their interaction with each other on the atomic center (A2b) and with external fields (A2c).

$$\hat{H} = \hat{H}_0 + \hat{H}_{int} + \hat{H}_{ex}, \tag{A1}$$

$$H_0 = \sum_{\sigma=+,-}\sum_k \gamma_k \left(a_{k\sigma}^+ a_{k\sigma} + b_{k\sigma}^+ b_{k\sigma}\right) + \sum_{\sigma=+,-}\sum_k t_k \left(a_{k\sigma}^+ b_{k\sigma} + b_{k\sigma}^+ a_{k\sigma}\right), \tag{A2a}$$

$$H_{int} = -J\sum_{j=1}^{N_0}(\hat{S}_{aj}^2 + \hat{S}_{bj}^2) \equiv -2J\sum_{j=1}^{N_0}(\hat{F}_j^2 + \hat{L}_j^2), \tag{A2b}$$

$$H_{ex} = g\mu_B \vec{H}_0 \sum_j (\vec{S}_{aj} + \vec{S}_{bj}) + g\mu_B \sum_j (\vec{H}_a \vec{S}_{aj} + \vec{H}_b \vec{S}_{bj}) \equiv$$
$$\equiv 2g\mu_B \vec{H}_0 \sum_j \hat{\vec{F}}_j + 2g\mu_B \vec{H}_Q \sum_j \hat{\vec{L}}_j, \qquad (A2c)$$

$$\hat{\vec{F}}_j = \left(\frac{\hat{S}_{aj} + \hat{S}_{bj}}{2}\right), \hat{\vec{L}}_j = \left(\frac{\hat{S}_{aj} - \hat{S}_{bj}}{2}\right), \qquad (A2d)$$

$$\vec{H}_a = -\vec{H}_b = \vec{H}_Q,$$

where $J$ is effective intraatomic exchange integral; $\hat{S}_{aj}, \hat{S}_{bj}$ are spin operators of one-electron atoms in $a,b$ positions expressed by the Fourier components of the Fermi operators $a_j^+(a_j), b_j^+(b_j)$ of creation (annihilation) of electrons in the cell $j$; $N_0$ is the number of cells; $\gamma_k, t_k$ are the Fourier components of hopping integrals in ($aa$, $bb$) and between ($ab$) planes formed by $a$ and $b$ atoms.

In the static limit, when using Hubbard-Stratonovich transformation and saddle-point approximation, the expression for the free energy per one state $\mathbf{F}(\xi,\eta) = \mathbf{F}(\xi,\eta)/2N_0$ at $T \to 0$ has the form [4]

$$\mathbf{F}(\xi,\eta) = E(\xi,\eta) + J(\xi - h)^2 + J \cdot (\eta - h_Q)^2 \qquad (A3)$$

$$E(\xi,\eta) = \frac{V_0}{16\pi^3} \sum_{m=1}^{4} \int d^3k \left[\{E^m(k,\xi,\eta)\Theta(\mu - E^m(k,\xi,\eta))\}\right], \qquad (A4)$$

where $\Theta(x)$ is Heaviside step-function; $\vec{\xi}, \vec{\eta}$ are exchange fields conjugate to $\hat{\vec{F}}$ and $\hat{\vec{L}}$ correspondingly; $E^m(k,\xi,\eta)$ are branches of the energy spectrum determined by solving the secular equation

$$\begin{vmatrix} \gamma_k - E & J(\xi_x + i\xi_y + \eta) & t_k^* & 0 \\ J(\xi_x - i\xi_y + \eta) & \gamma_k - E & 0 & t_k^* \\ t_k & 0 & \gamma_k - E & J(\xi_x + i\xi_y - \eta) \\ 0 & t_k & J(\xi_x - i\xi_y - \eta) & \gamma_k - E \end{vmatrix} = 0$$

$$E^{1,2}(k,\xi,\eta) = \gamma_k \pm \sqrt{t_k^2 + (J)^2(\vec{\eta}^2 + \vec{\xi}^2) + 2J\sqrt{t_k^2\vec{\xi}^2 + (J)^2(\vec{\xi}\vec{\eta})^2}}$$

$$E^{3,4}(k,\xi,\eta) = \gamma_k \pm \sqrt{t_k^2 + (J)^2(\vec{\eta}^2 + \vec{\xi}^2) - 2J\sqrt{t_k^2\vec{\xi}^2 + (J)^2(\vec{\xi}\vec{\eta})^2}} \qquad (A5)$$

Exchange fields $\vec{\xi}, \vec{\eta}$ for given values of the direction cosines $e_x$, $e_y$ are connected with the mean values of ferromagnetic $m_0 = g\langle\hat{F}\rangle\mu_B$ and antiferromagnetic $m_Q = g\langle\hat{L}\rangle\mu_B$ components of the atomic magnetic moment (directed along corresponding fields $H_0, H_Q$) by the relations derived

from conditions

$$m_0 = (\xi - h)g\mu_B, \qquad (A6a)$$

$$m_Q = (\eta - h_Q)g\mu_B, \qquad (A6b)$$

Where $h = \dfrac{g\mu_B H_0}{2J}$, $h_Q = \dfrac{g\mu_B H_Q}{2J}$.

Assuming that the electron spectrum in the nearest neighbor approximation admits a factorization by $\vec{k}$ (for example $t_k$ depends on $k_z$, and $\gamma_k$ only on $k_x, k_y$), we can pass from the three-dimensional integration over $\vec{k}$ to the "two-dimensional" one, introducing the functions

$$g_{00}(\varepsilon) = \left(\frac{V_0}{8\pi^3}\right)^{2/3} \int dk_x dk_y \delta(\varepsilon - \gamma_k), \qquad (A7a)$$

$$g_{11}(\varepsilon_1) = \left(\frac{V_0}{8\pi^3}\right)^{1/3} \int dk_z \delta(\varepsilon_1 - t_k) \qquad (A7b)$$

Then for (A4), we obtain

$$E(\xi,\eta) = \frac{1}{2}\sum_m \int d\varepsilon\, d\varepsilon_1\, E^m(\varepsilon,\varepsilon_1,\xi,\eta)\Theta(\mu - E^m(\varepsilon,\varepsilon_1,\xi,\eta))g_0(\varepsilon)g_1(\varepsilon_1) \qquad (A8)$$

where the branches of the electronic spectrum $E^m(\varepsilon,\varepsilon_1,\xi,\eta)$ can be written as

$$E^{1,2}(\varepsilon,\varepsilon_1,\xi,\eta) = \varepsilon \pm \sqrt{\varepsilon_1^2 + (J)^2(\vec{\eta}^2 + \vec{\xi}^2) + 2J\sqrt{\varepsilon_1^2\vec{\xi}^2 + (J)^2(\vec{\xi}\vec{\eta})^2}} \qquad (A9a)$$

$$E^{3,4}(\varepsilon,\varepsilon_1,\xi,\eta) = \varepsilon \pm \sqrt{\varepsilon_1^2 + (J)^2(\vec{\eta}^2 + \vec{\xi}^2) - 2J\sqrt{\varepsilon_1^2\vec{\xi}^2 + (J)^2(\vec{\xi}\vec{\eta})^2}} \qquad (A9b)$$

$$g_0(\varepsilon) = g_{00}(\varepsilon)/\int g_{00}(\varepsilon)d\varepsilon, \quad g_1(\varepsilon_1) = g_{11}(\varepsilon_1)/\int g_{00}(\varepsilon_1)d\varepsilon_1 \qquad (A10)$$

State equation $\partial F/\partial \xi = 0$ (A11a), $\partial F/\partial \eta = 0$ (A11b) for $H_Q = 0$ supplemented by the equations for the chemical potential (A11c) have the form

$$h = \xi + \frac{1}{2J}\frac{\partial E(\xi,\eta)}{\partial \xi} \qquad (A11a)$$

$$0 = \eta + \frac{1}{2J}\frac{\partial E(\xi,\eta)}{\partial \eta} \qquad (A11b)$$

$$n = \frac{1}{2}\int g_1(\varepsilon_1)g_0(\varepsilon)\cdot \sum_m [\Theta(\mu - E_m(\varepsilon,\varepsilon_1,\xi,\eta))]d\varepsilon\, d\varepsilon_1 = \frac{1}{2}\int_{}^{\mu} G(E)dE \qquad (A11c)$$

The relation between the model density of electronic states $G(\varepsilon)$ and functions $g_0(\varepsilon)$, $g_1(\varepsilon_1)$ is determined by the expression

$$G(E) = \sum_{m=1}^{4} \int d\varepsilon \, d\varepsilon_1 \delta(E - E_m(\varepsilon, \varepsilon_1, \xi, \eta)) g_0(\varepsilon) g_1(\varepsilon_1) \tag{A12}$$

# Figure Captions

**Fig. 1**. (color online) Magnetic phase diagram of the $Fe_{2-x}Mn_xAs$ [2].

**Fig. 2**. (color online) Manganese content dependences of experimental M, μ and calculated from first principles magnetic moments $M_{FM}$, $M_{FIM-I}$ of the $Fe_{2-x}Mn_xAs$ and $Fe_{2.1-x}Mn_xAs$ systems. (a) values of M are taken from the data of [2]; (b) values of M are taken from the data of [1] and the measurements of field dependences of magnetization in [6]; (c) $μ_{PM}$ corresponds to paramagnetic moment.

**Fig. 3**. (color online) Field dependence of the magnetization of the monocrystalline $Fe_{0.786}Mn_{1.414}As$ ($a = 2.2$) sample perpendicular (a) and along (b) tetragonal axis $c$ under uniaxial pressure. 1,4-initial low-ferrimagnetic state; 2,3,5,6 - antiferromagnetic state, (a) uniaxial pressure is perpendicular to the tetragonal axis, (b) uniaxial pressure is parallel to the tetragonal axis. P, kbar: 1,3,4,5-0.001; 2-0.52; 6-0.48. T, K: 1-321; 2,3-343, 4-313, 5,6–331 [8].

**Fig. 4**. (color online) Magnetic P-T phase diagram for the nonstoichiometric samples for different pressures. 1, I - P⊥c, 2, II - hydrostatic pressure, 3, III - P∥c. Arabic numerals – $a = 2.2$, $x = 1.414$; roman – $a = 1.95$, $x = 1.19$. Neel temperature ($T_N$) correspond to atmospheric pressure [8].

**Fig. 5**. (color online) d-electron density of states of some $Fe_{2-x}Mn_xAs$ compounds in nonmagnetic phase reduced to common Fermi level. The characteristic values of DOS(E) are marked by $E_i$, $D_j$.

**Fig. 6**. (color online) Evolution of the electronic filling $N_d$ and some characteristics of the shape of density of electronic states $DOS_{dNM}$ at varied manganese content.

**Fig. 7**. (color online) Evolution of the electronic filing and shape characteristics of density of electronic states $DOS_{dNM}$ at varied relative volume. (a) hydrostatic compression($c/a$) = const; (b) uniaxial compression along the $c$ axis($Δc<0$, $Δa>0$); (c) uniaxial compression perpendicular to the $c$ axis($Δc>0$, $Δa<0$).

**Fig. 8**. (color online) The bare functions $g_{00}(ε)$, $g_{11}(ε_1)$ and model density of electronic states.

Vertical line indicates the Fermi level at $n = 1.2128$ ($x = 1.31$).

**Fig. 9**. (color online) The shape characteristics of the model density of electronic states $G_{NM}(E)$, electronic filing $n$, and the angle $\Theta$ at varied content of Mn.

**Fig. 10**. (color online) Magnetic moment dependence of the free energy of the canted **F2**$(m_0)$ and collinear ferromagnetic **F1**$(m_0)$ states for varied electronic filling $n(x)$. The values $m_{00}$ and $m_{FM0}$ correspond to the equilibrium values of the components of magnetic moment at $B = 0$. The inset shows how the equilibrium values of vectors of ferromagnetism $m_{00}$ and antiferromagnetism $m_{Q0}$ and angle $\Theta$ between them are related to decreasing electronic filling $n$ of d-state (increasing manganese content $x$).

**Fig. 11**. (color online) Model magnetization curves for varied electronic filling $n$. (a) Simulation of reversible magnetic field induced transitions AF-LFi$_1$-LFi$_2$; dependence $m_Q(B)$ describes field changes for antiferromagnetic components which accompany the changes of the ferromagnetic components $m_0(B)$ for the corresponding value $n$; the values $5m_{Fm}(n)$ are compared with calculated from first principles magnetic moments of $M_{FM}(x)$ per an ion in the FM phase (Fig. 2c); (b) curve 1 can be compared with irreversible transition AF-LFi$_1$ in Fe$_{0.815}$Mn$_{1.185}$As [5] (solid line), curves 2, 3 are compared with reversible LFi$_1$-LFi$_2$ transition in Fe$_{0.71}$Mn$_{1.29}$As, Fe$_{0.65}$Mn$_{1.35}$As [6, 9] (solid line) curve 5 simulates the transitions of the second (AF-LFi$_1$) and first (LFi$_1$-LFi$_2$) order typical for the samples with a high content of manganese. Experimental field plots 1 were measured in [5] at T=4.2K, 2 and 3 in [6] at T =13K and 77K.

**Fig. 12**.(color online) The unit cell of the model at non-orthogonal arrangement of vectors of ferromagnetism $\xi = 2m_0$ and antiferromagnetism $\eta = 2m_Q$.

**Fig. 13**. (color online) Model magnetization curves in normal (empty symbols) and deformed (crossed symbols) states at $n = 1.2128$ ($x = 1.31$). (b,c) the curves are compared with curves 6,2 in Fig. 3.

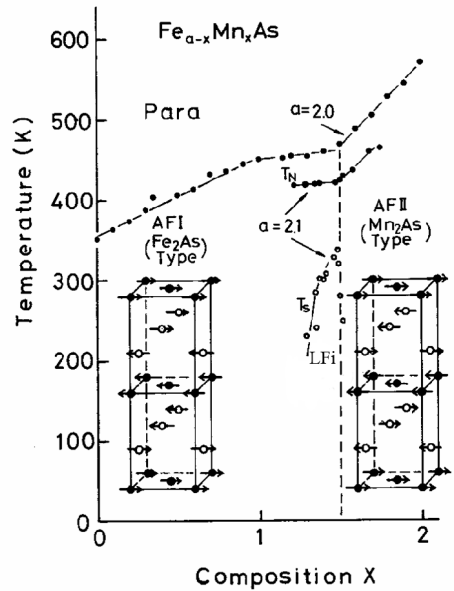

**Fig. 1**. (color online) Magnetic phase diagram of the $Fe_{2-x}Mn_xAs$ [2].

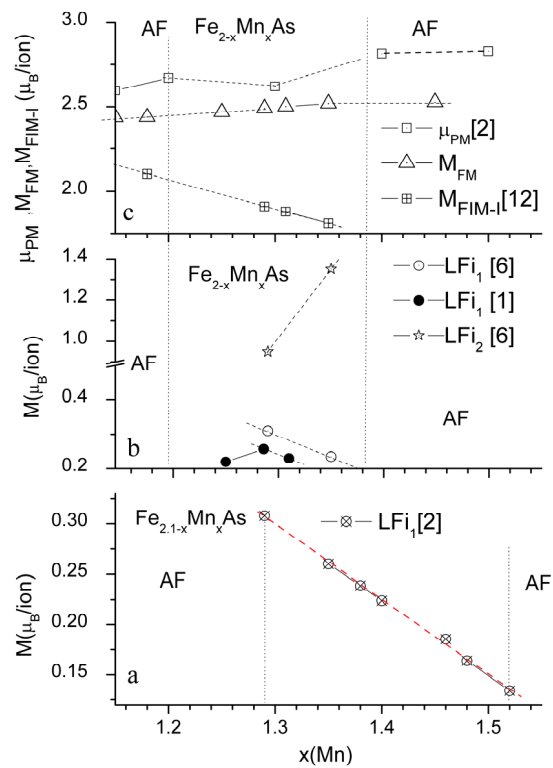

**Fig. 2**. (color online) Manganese content dependences of experimental M, μ and calculated from first principles magnetic moments $M_{FM}$, $M_{FIM-I}$ of the $Fe_{2-x}Mn_xAs$ and $Fe_{2.1-x}Mn_xAs$ systems. (a) values of M are taken from the data of [2]; (b) values of M are taken from the data of [1] and the measurements of field dependences of magnetization in [6]; (c) $\mu_{PM}$ corresponds to paramagnetic moment.

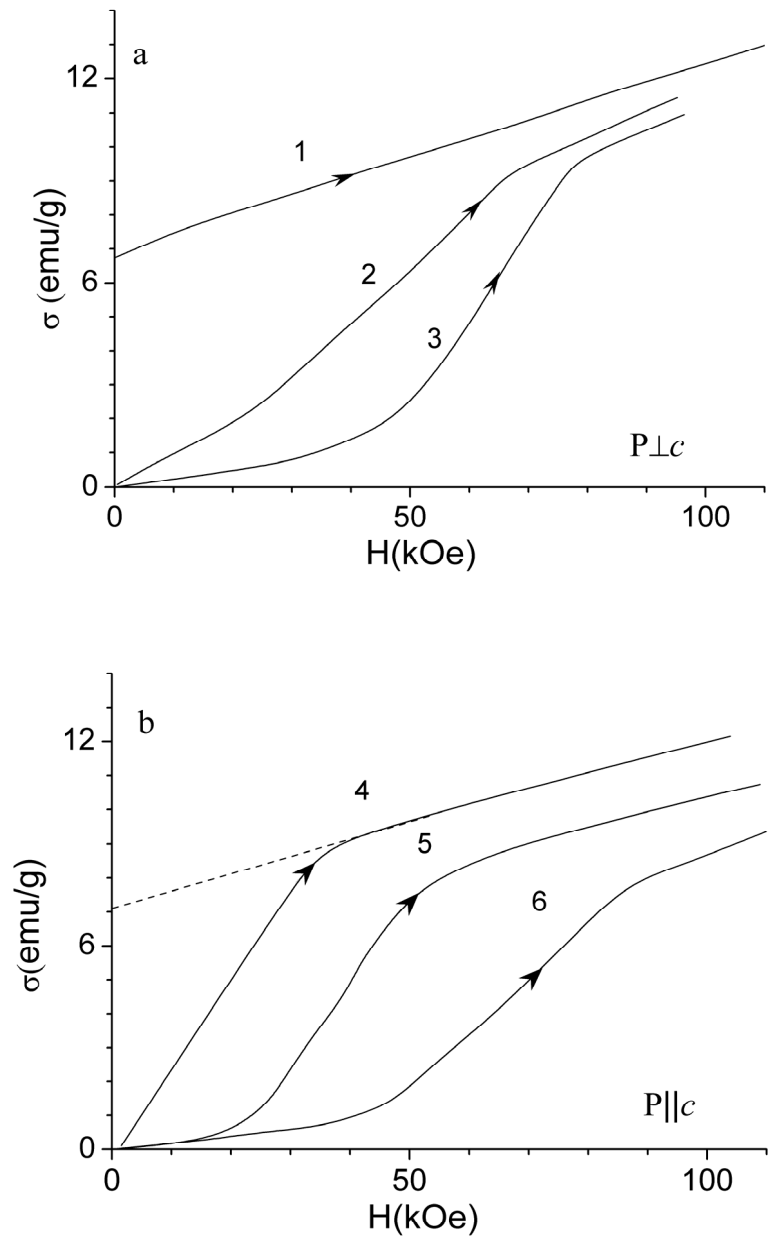

**Fig. 3**. (color online) Field dependence of the magnetization of the monocrystalline $Fe_{0.786}Mn_{1.414}As$ ($a = 2.2$) sample perpendicular (a) and along (b) tetragonal axis $c$ under uniaxial pressure. 1,4-initial low-ferrimagnetic state; 2,3,5,6 - antiferromagnetic state, (a) uniaxial pressure is perpendicular to the tetragonal axis, (b) uniaxial pressure is parallel to the tetragonal axis. P, kbar: 1,3,4,5-0.001; 2-0.52; 6-0.48. T, K: 1-321; 2,3-343, 4-313, 5,6–331 [8].

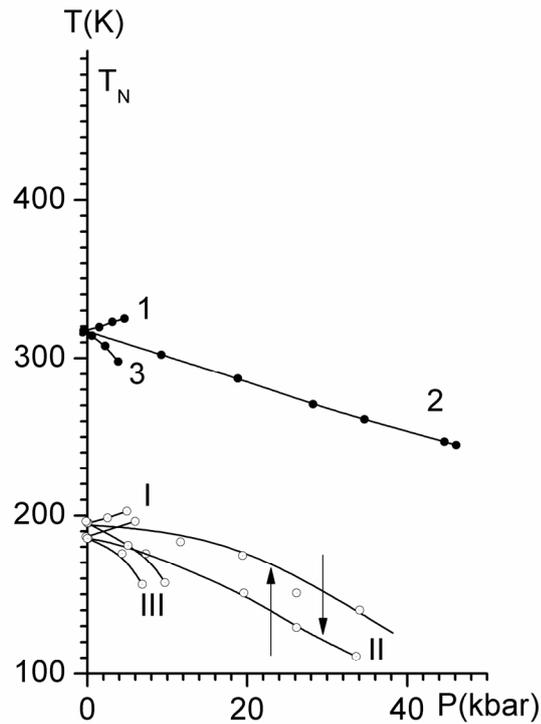

**Fig. 4**. (color online) Magnetic P-T phase diagram for the nonstoichiometric samples for different pressures. 1, I - P⊥c, 2, II - hydrostatic pressure, 3, III - P∥c. Arabic numerals – $a = 2.2$, $x = 1.414$; roman – $a = 1.95$, $x = 1.19$. Neel temperature ($T_N$) correspond to atmospheric pressure [8].

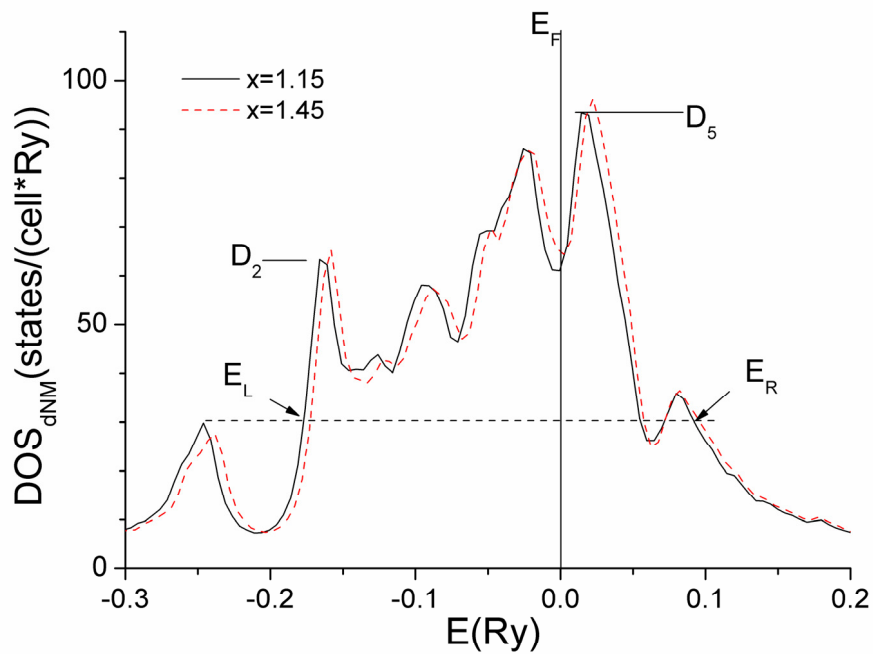

**Fig. 5**. (color online) d-electron density of states of some $Fe_{2-x}Mn_xAs$ compounds in nonmagnetic phase reduced to common Fermi level. The characteristic values of DOS(E) are marked by $E_i$, $D_j$.

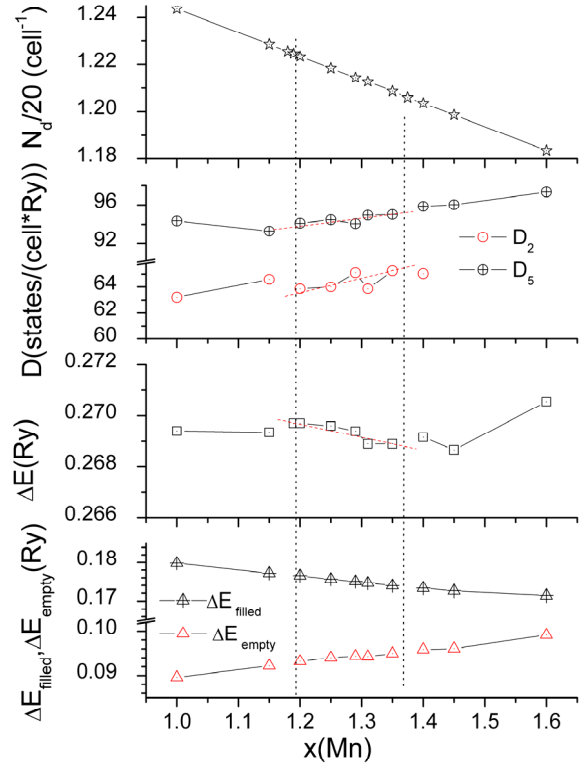

**Fig. 6**. (color online) Evolution of the electronic filling $N_d$ and some characteristics of the shape of density of electronic states $DOS_{dNM}$ at varied manganese content.

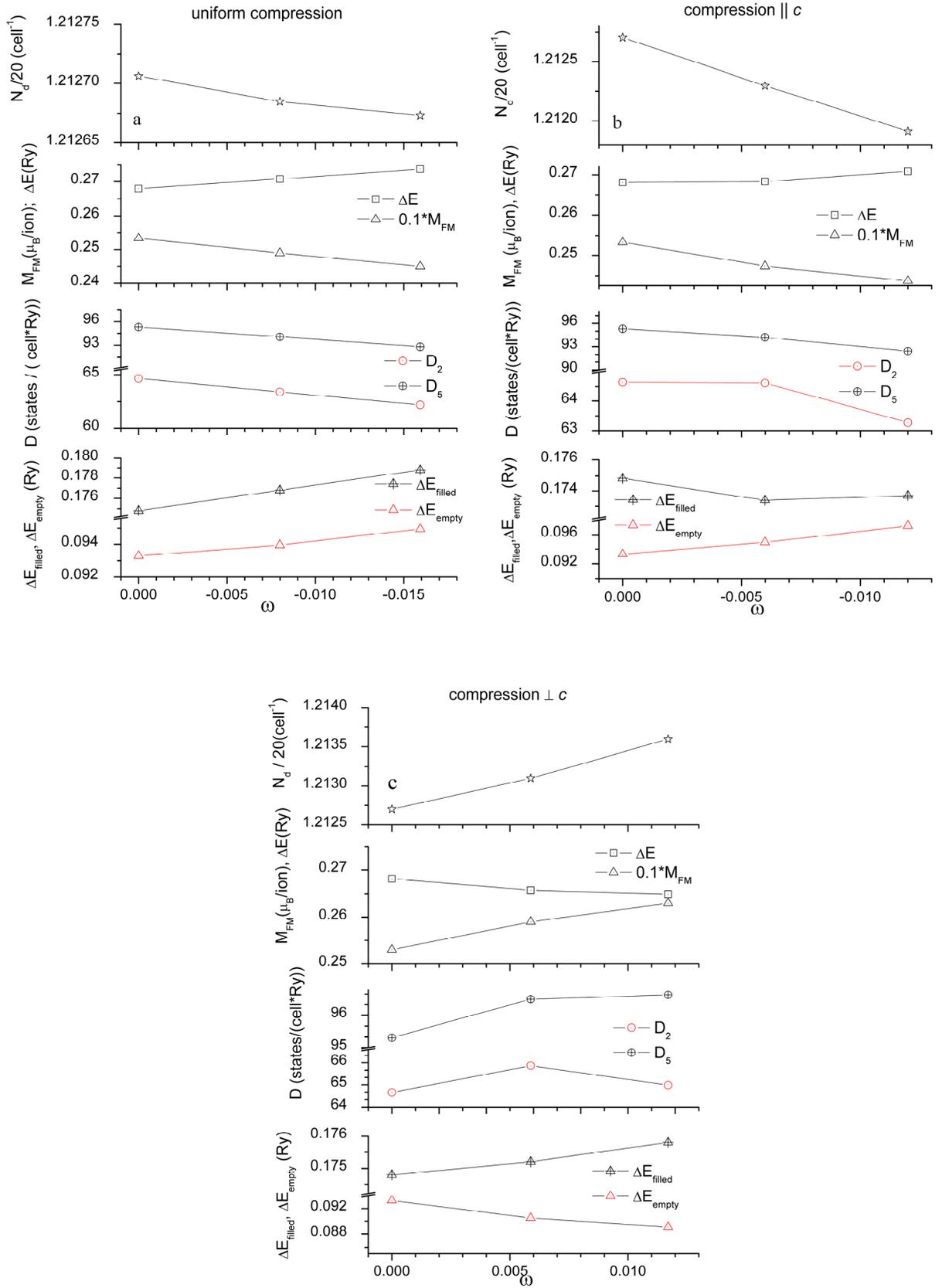

**Fig. 7**. (color online) Evolution of the electronic filing and shape characteristics of density of electronic states $DOS_{dNM}$ at varied relative volume. (a) hydrostatic compression($c/a$) = const; (b) uniaxial compression along the $c$ axis($\Delta c<0$, $\Delta a>0$); (c) uniaxial compression perpendicular to the $c$ axis($\Delta c>0$, $\Delta a<0$).

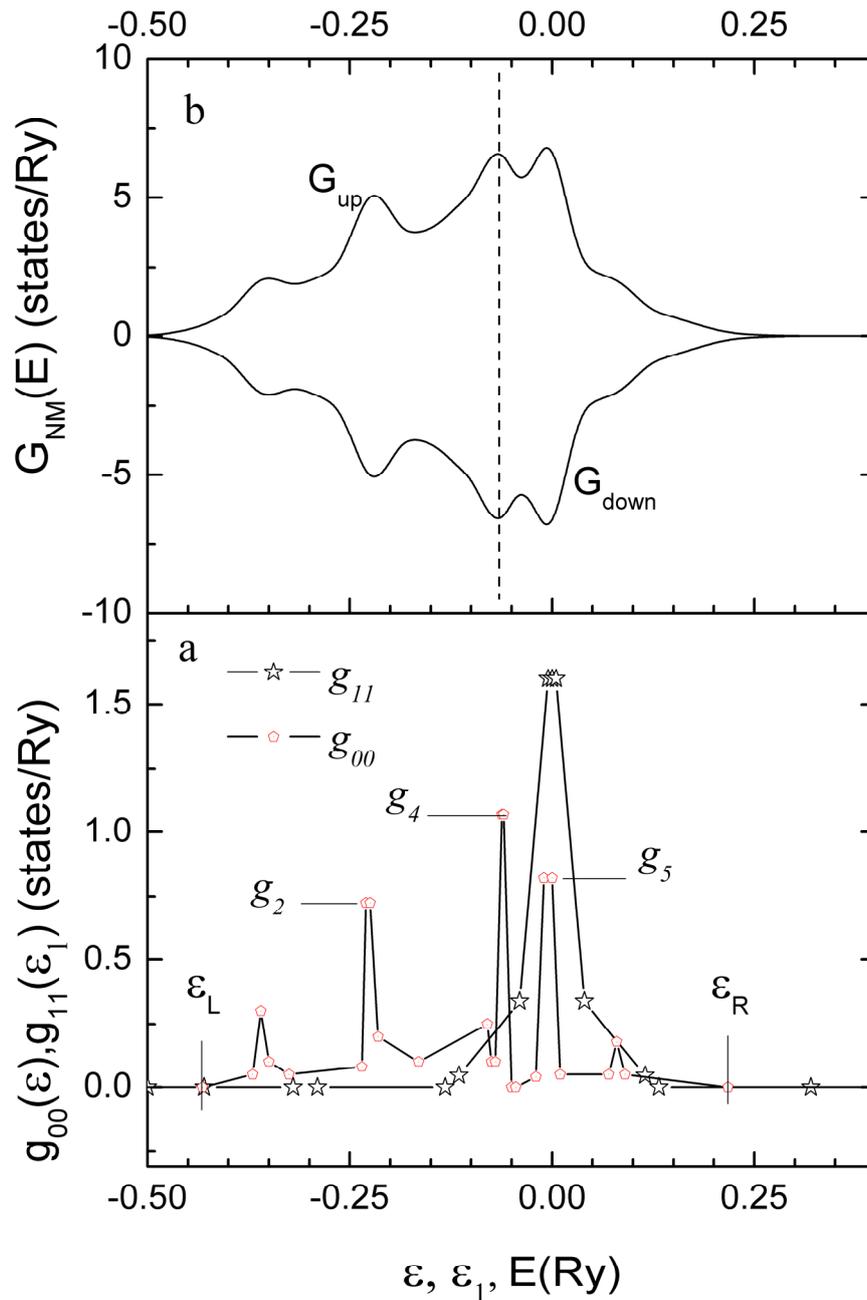

**Fig. 8**. (color online) The bare functions $g_{00}(\varepsilon)$, $g_{11}(\varepsilon_1)$ and model density of electronic states. Vertical line indicates the Fermi level at $n = 1.2128$ ($x = 1.31$).

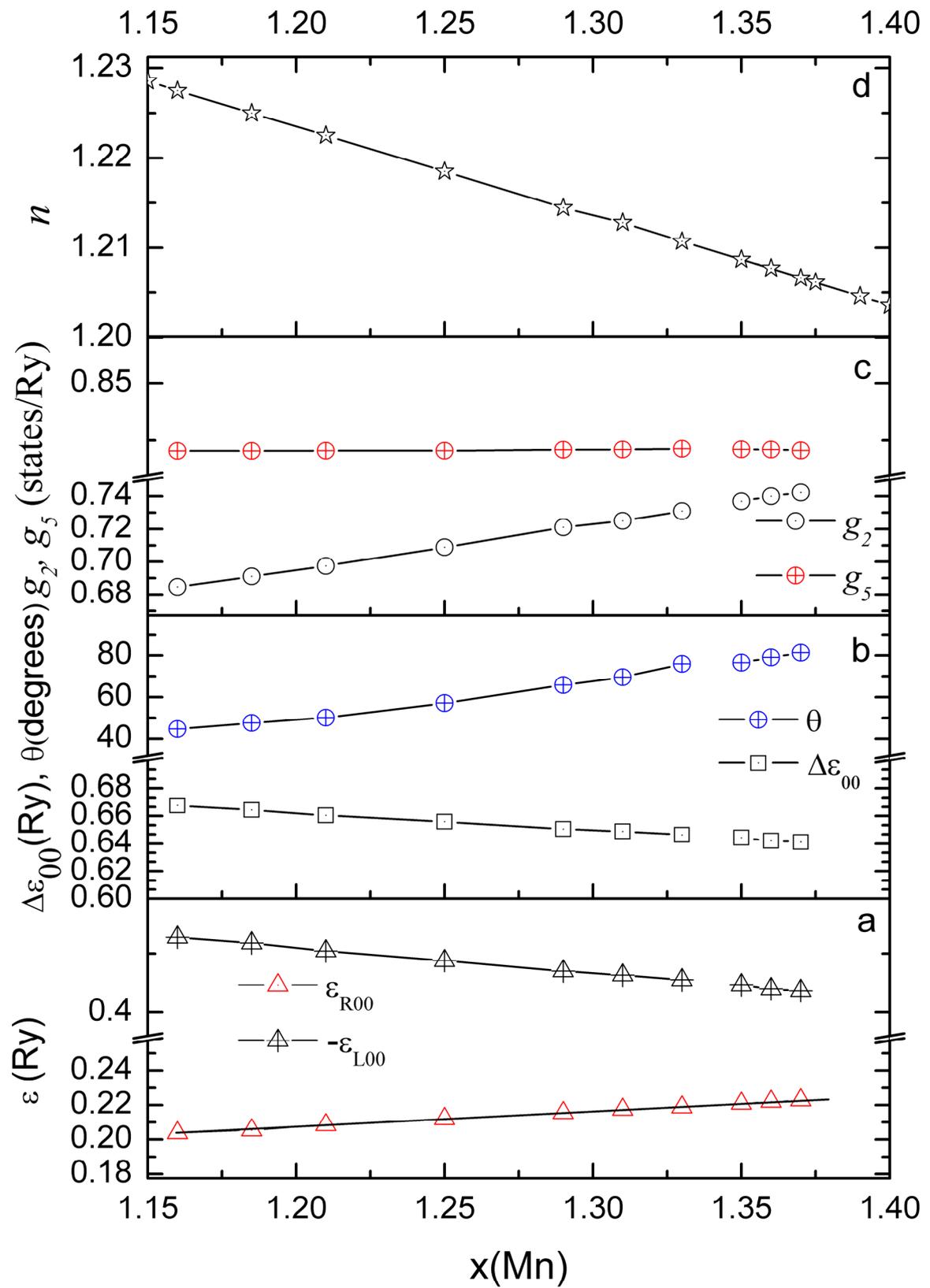
**Fig. 9.** (color online) The shape characteristics of the model density of electronic states $G_{NM}(E)$, electronic filing $n$, and the angle $\Theta$ at varied content of Mn.

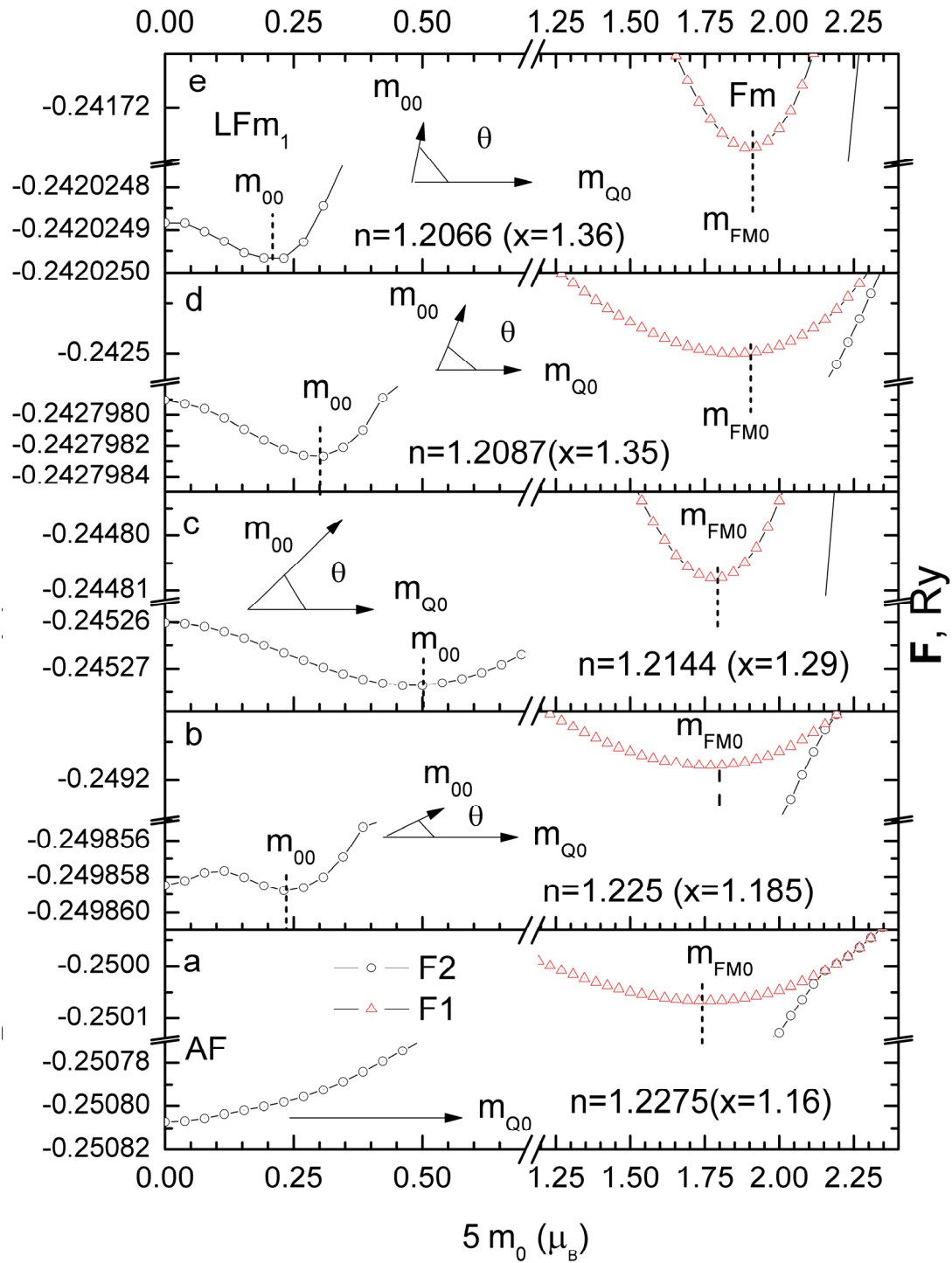

**Fig. 10**. (color online) Magnetic moment dependence of the free energy of the canted **F2**($m_0$) and collinear ferromagnetic **F1**($m_0$) states for varied electronic filling $n(x)$. The values $m_{00}$ and $m_{FM0}$ correspond to the equilibrium values of the components of magnetic moment at $B = 0$. The inset shows how the equilibrium values of vectors of ferromagnetism $m_{00}$ and antiferromagnetism $m_{Q0}$ and angle $\Theta$ between them are related to decreasing electronic filling $n$ of d-state (increasing manganese content $x$).

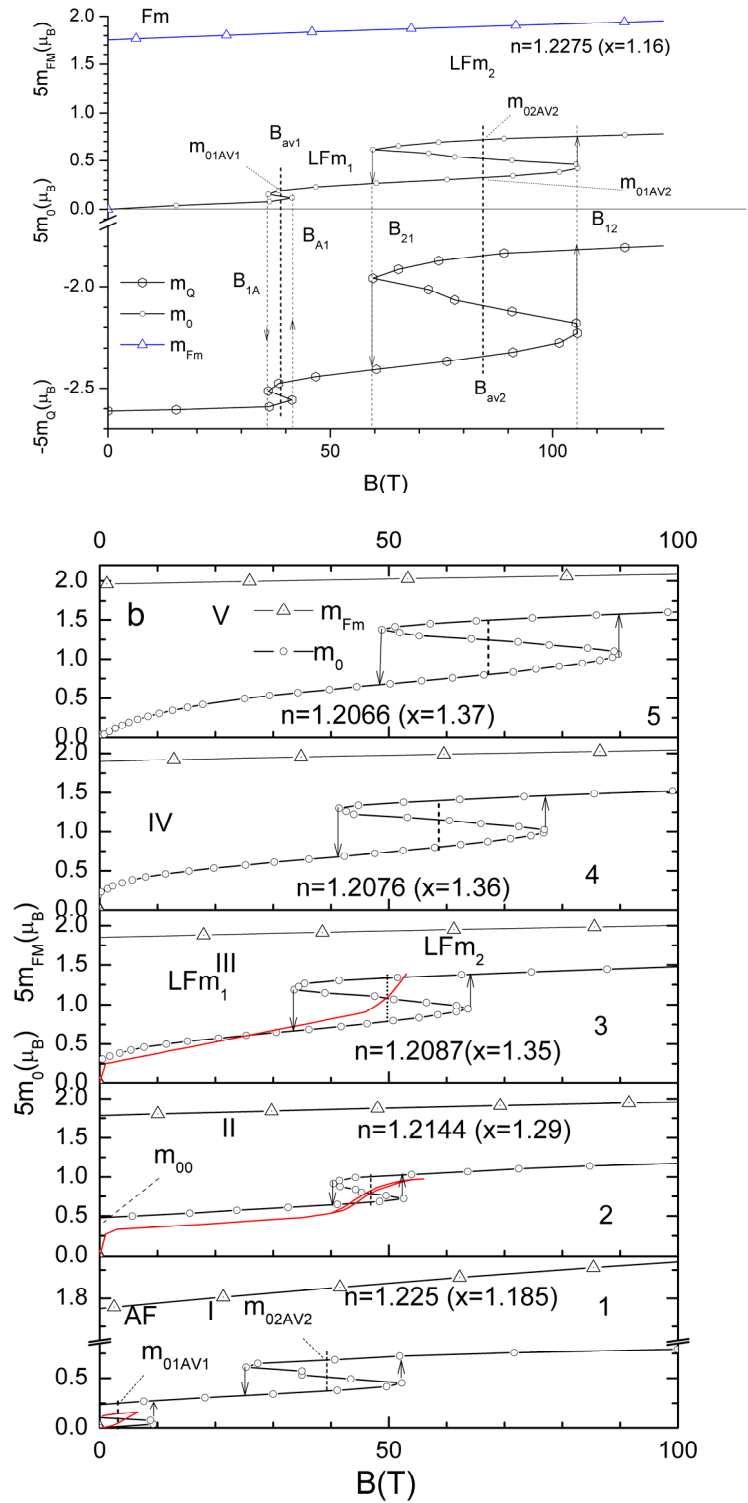

**Fig. 11**. (color online) Model magnetization curves for varied electronic filling *n*. (a) Simulation of reversible magnetic field induced transitions AF-LFi$_1$-LFi$_2$; dependence m$_Q$(B) describes field changes for antiferromagnetic components which accompany the changes of the ferromagnetic components m$_0$(B) for the corresponding value *n*; the values 5m$_{Fm}$(n) are compared with calculated from first principles magnetic moments of M$_{FM}$(x) per an ion in the FM phase (Fig. 2c); (b) curve 1 can be compared with irreversible transition AF-LFi$_1$ in Fe$_{0.815}$Mn$_{1.185}$As [5] (solid line), curves 2, 3 are compared with reversible LFi$_1$-LFi$_2$ transition in Fe$_{0.71}$Mn$_{1.29}$As, Fe$_{0.65}$Mn$_{1.35}$As [6, 9] (solid line) curve 5 simulates the transitions of the second (AF-LFi$_1$) and first (LFi$_1$-LFi$_2$) order typical for the samples with a high content of manganese. Experimental field plots 1 were measured in [5] at T=4.2K, 2 and 3 in [6] at T =13K and 77K.

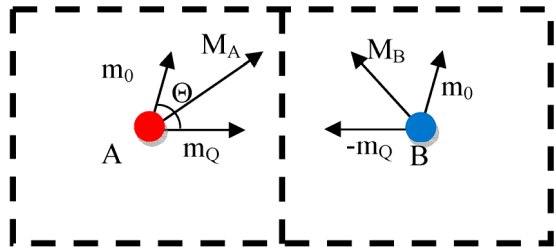

**Fig. 12.** (color online) The unit cell of the model at non-orthogonal arrangement of vectors of ferromagnetism ξ = 2m₀ and antiferromagnetism ŋ = 2m_Q.

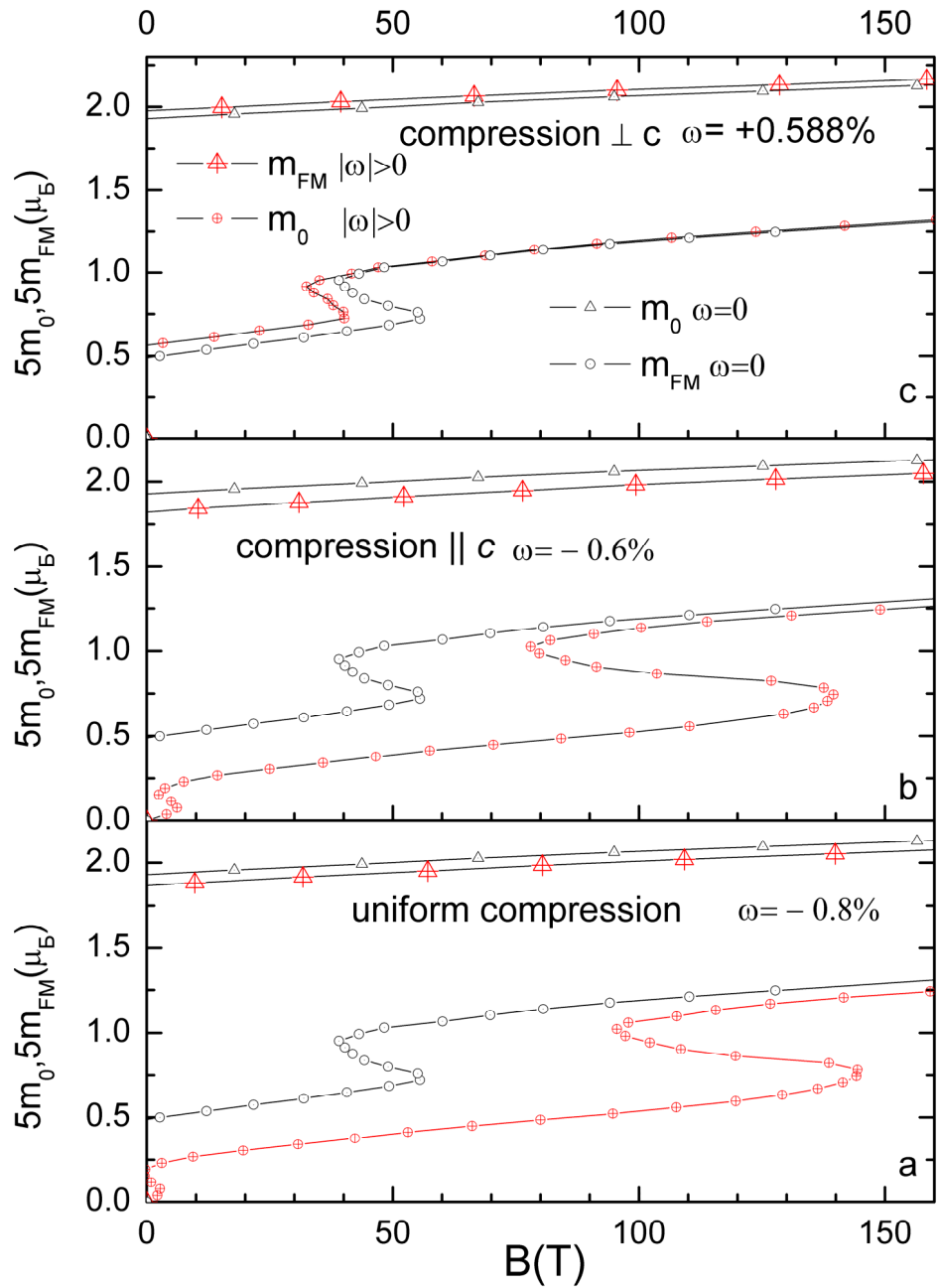

**Fig. 13.** (color online) Model magnetization curves in normal (empty symbols) and deformed (crossed symbols) states at $n = 1.2128$ ($x = 1.31$). (b,c) the curves are compared with curves 6,2 in Fig. 3.